\documentclass[aps,prl,reprint,twocolumn,superscriptaddress,showpacs,longbibliography]{revtex4-1}
\usepackage{graphicx,dcolumn,hyperref,braket,amsmath,amssymb,xcolor} 
\usepackage{epstopdf}
\usepackage{natbib}
\usepackage{mathtools}

\begin{document}

\title{Many-particle interference to test Born's rule}

\author{Marc-Oliver Pleinert}
\affiliation{Institut f\"{u}r Optik, Information und Photonik, \\ Friedrich-Alexander-Universit\"{a}t Erlangen-N\"{u}rnberg (FAU), 91058 Erlangen, Germany}
\affiliation{Erlangen Graduate School in Advanced Optical Technologies (SAOT), Friedrich-Alexander-Universit\"{a}t Erlangen-N\"{u}rnberg (FAU), 91052 Erlangen, Germany}
\author{Joachim von Zanthier}
\affiliation{Institut f\"{u}r Optik, Information und Photonik, \\ Friedrich-Alexander-Universit\"{a}t Erlangen-N\"{u}rnberg (FAU), 91058 Erlangen, Germany}
\affiliation{Erlangen Graduate School in Advanced Optical Technologies (SAOT), Friedrich-Alexander-Universit\"{a}t Erlangen-N\"{u}rnberg (FAU), 91052 Erlangen, Germany}
\author{Eric Lutz}
\affiliation{Institute for Theoretical Physics I, University of Stuttgart, D-70550 Stuttgart, Germany}

\begin{abstract}
	Born's rule, one of the cornerstones of quantum mechanics, relates detection probabilities to the modulus square of the wave function.  Single-particle interference is accordingly limited to pairs of quantum paths and higher-order interferences  are prohibited. Deviations from Born's law have been quantified via the Sorkin parameter which is proportional to the third-order term. We here extend this formalism to many-particle interferences and find that they exhibit a much richer structure. We demonstrate, in particular, that all interference terms of order $(2M+1)$ and greater vanish for $M$~particles. We further introduce a family of many-particle Sorkin parameters and show that they are exponentially more sensitive  to deviations from Born's rule than their single-particle counterpart.
\end{abstract}

\maketitle

According to Born's rule the probability of a  measurement outcome is given by the modulus square of the corresponding probability amplitude~\cite{Born:1926}. This fundamental postulate of quantum mechanics establishes a link between the (deterministic) mathematical formalism and experiment. It additionally introduces a random component into the theory. In view of its significance, many attempts to replace the postulate by a derivation from underlying principles have been made~\cite{Everett:1957,Hartle:1968,Farhi:1989,Deutsch:1999,Zurek:2003,Saunders:2004}.
However, to date none of these derivations seems to be generally accepted~\cite{Squires:1990,Cassinello:1996,Barnum:2000,Schlosshauer:2005}. Violations of the rule have furthermore been predicted, for instance in cosmology~\cite{Page:2009,Aguirre:2011} and black hole physics~\cite{Marolf:2016}. Experimental tests of Born's law are therefore crucial to assess its range of validity.

A direct consequence of Born's rule is that single-particle quantum interference originates only from pairs of quantum paths~\cite{Sorkin:1994}. Interferences of higher order than the second therefore do not occur  in quantum mechanics for single particles.
The vanishing of third-order interference has been investigated in the context of a general quantum measure theory by Sorkin~\cite{Sorkin:1994}.  This has led to the introduction of the so-called Sorkin parameter $\kappa$, Eq.~\eqref{eq:Sorkin-parameter} below, which vanishes if Born's rule holds. The parameter $\kappa$ has been measured in  three and five-slit experiments  with single photons~\cite{Sinha:2010,Sollner:2012,Kauten:2017} and single molecules~\cite{Cotter:2017}. It has been found to   be smaller than $3\times 10^{-5}$ in the classical light regime and $2\times 10^{-3}$ in the quantum regime~\cite{Kauten:2017}. These findings rule out  multi-order  single-particle interference~\cite{Lee:2017} and confirm Born's law to that level of precision.

Quantum mechanics, however, is not limited to single-particle interference phenomena. It also allows for {many-particle} interference in the case of indistinguishable particles. For example, two photons, albeit non-interacting, can influence each other via two-particle interference as in the Hong-Ou-Mandel experiment~\cite{Hong:1987}. Here, two indistinguishable photons impinging on a beam splitter will not exit the output ports of the beam splitter separately due to the destructive interference of the two different {two-photon} paths, where both photons are transmitted or reflected, respectively.
Many-particle interference is mathematically  richer and physically more subtle than single-particle interference~\cite{Pan:2012,Tichy:2014}. Recently, two groups have independently observed genuine three-photon interference that does not originate from two- or single-photon interference~\cite{Agne:2017,Menssen:2017}. 
Many-particle interference is not only of interest from a fundamental point of view~\cite{Hong:1987,Tichy:2014,Agne:2017,Menssen:2017,Pan:2012,Mahrlein:2017}, but has also been exploited in imaging ~\cite{Hanbury-Brown:1956,Thiel:2007,Schneider:2017}, metrology~\cite{Giovannetti:2011,Su:2017} and for quantum information processing~\cite{Aaronson:2011,Wang:2017}. Multi-boson interference in unitary networks has for instance been shown to be computationally hard, yielding the promising model of boson sampling~\cite{Aaronson:2011}. 

In this paper, we propose to use many-particle interference for a fundamental test of Born's rule, and thus of quantum theory.
To lay out the idea, we first recapitulate the concept of single-particle interference in triple-slit experiments and introduce the  Sorkin parameter. We then study more general setups with greater number of slits, where many-particle interference, and hence higher-order correlations among the particles, can be observed. 
In this setting, we generalize Sorkin's idea to incorporate many-particle interference displaying a richer structure and far more nonzero higher-order interference terms than single-particle interference. 
We derive higher-order sum rules that can be used to test Born's rule and introduce a family of many-particle Sorkin parameters. We demonstrate, in particular, that all interference terms of order $(2M+1)$ and larger vanish for $M$ particles. We finally show that the generalized Sorkin parameters are exponentially more sensitive with increasing $M$ to deviations from Born's rule than their single-particle counterpart.

\paragraph*{Single-particle interference.}
In the standard double-slit experiment, interference may be quantified by comparing the pattern $P_{AB}(\mathbf{r},t)$ observed when both slits $A$ and $B$ are open to the classical expectation of the sum of two single slits, $P_A(\mathbf{r},t) + P_B(\mathbf{r},t)$, where either slit $A$ or slit $B$ is open~\cite{Feynman:1965}.
Here, $P(\mathbf{r},t)$ stands for the detection probability of a single particle at position $\mathbf{r}$ and time $t$. 
The interference term is thus given by
\begin{eqnarray}\label{eq:I1_AB}
I^{(1)} &:=& P_{AB} - ( P_A + P_B ) \, ,
\end{eqnarray}
where we indicate in the superscript that we are dealing with single-particle interference (and omit the arguments $(\mathbf{r},t)$ for simplicity).
Equation \eqref{eq:I1_AB} can  be computed quantum-mechanically by applying Born's rule (BR) and the superposition principle (SP):\begin{eqnarray}\label{eq:G1AB-derivation}
P_{AB} &\overset{(BR)}{=}& \left| \psi_{AB}\right|^2 \overset{(SP)}{=} \left| \psi_{A} + \psi_{B} \right|^2  \\
&=& \left| \psi_{A} \right|^2 + \left| \psi_{B} \right|^2 + \psi_{A}^* \psi_{B} + \psi_{B}^* \psi_{A}  \notag \\
&=& P_A + P_B + I^{(1)} \notag \, ,
\end{eqnarray}
where $\psi_{X}$ denotes the wave function of a particle at the detection plane when slit $X$ (slits $XY$) is (are) open.
Any nonzero value of $I^{(1)}$ indicates a deviation from the ordinary (incoherent) sum of the subsystems $A$ and $B$, and thus any interference present between those. 
In experiments with particles whose wave functions are spatially coherent over the slits,
$I^{(1)}$ typically shows a sinusoidal behavior~\cite{Feynman:1965}.

Sorkin has generalized this approach to a hierarchy of interferences  and defined quantum mechanics as a measure theory fulfilling a higher-order interference sum-rule than classical mechanics~\cite{Sorkin:1994}. 
In classical physics (with distinguishable particles), the interference term $I^{(1)}$ is zero, yielding the classical additivity of mutually exclusive events. By contrast, the formalism of quantum mechanics predicts a non-zero {second-order} interference term $I^{(1)} \equiv I^{(1)}_2 \neq 0$, while all higher-order interference terms $I^{(1)}_N$ $(N\geq3)$ vanish~\cite{Sorkin:1994}.
The usual interference term $I^{(1)}_2$  emerges naturally in double-slit experiments and marks off classical from quantum mechanics. On the other hand,  the third-order interference term 
$I^{(1)}_3 := P_{ABC} - (P_{AB} + P_{AC} + P_{BC}) + ( P_A + P_B + P_C)$
turning up in triple-slit setups allows one to distinguish quantum mechanics from generalized probabilistic theories. 
The pattern $P_{ABC}$ observed after a triple-slit can again be derived by applying BR and SP~\cite{Sorkin:1994}.
It turns out that all occurring interference terms are already included in the double slit terms $P_{XY}$. The triple slit pattern can hence be written as a sum of the comprised double and single slits and no genuine third-order interference term survives,  $I_3^{(1)}=0$. This is commonly referred to as the nonexistence of multi-order single-particle interference~\cite{Sorkin:1994,Sinha:2010}.
The Sorkin parameter is then defined as the third-order interference term normalized  to the central maximum of the triple slit,
\begin{eqnarray} \label{eq:Sorkin-parameter}
\kappa(\mathbf{r})=\frac{I_3^{(1)}(\mathbf{r})}{P_{ABC}(\mathbf{r}'=0)} \, .
\end{eqnarray} 
Equation \eqref{eq:Sorkin-parameter} can be used to quantify deviations from Born's rule  which predicts $\kappa=0$.

\paragraph*{Many-particle interference.}

\begin{figure*}
	\centering \includegraphics[width=\textwidth]{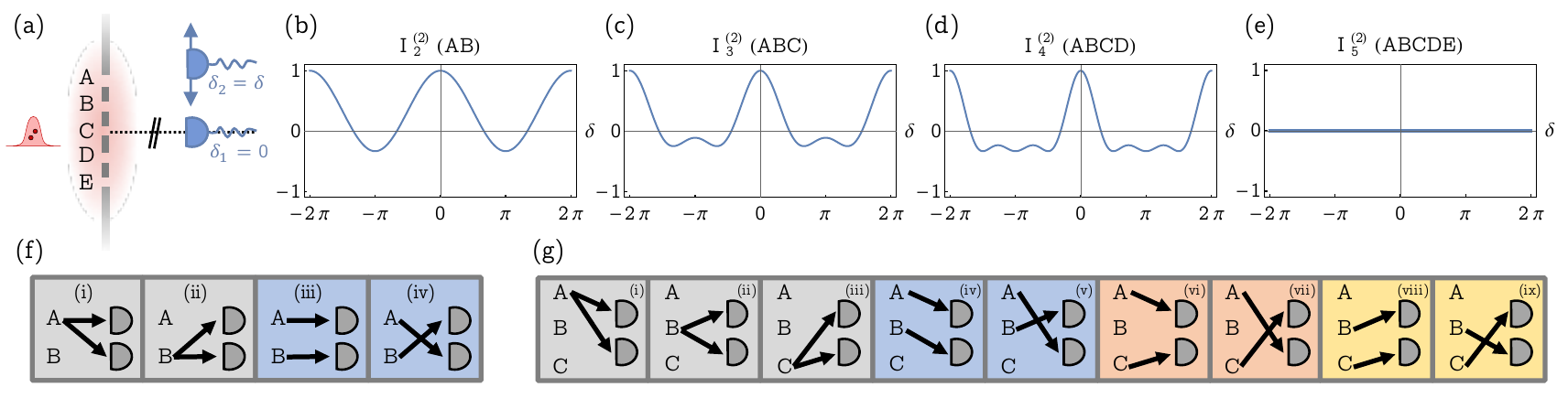}
	\caption{\label{fig:many-particle-interference} Two-particle interference in slit-experiments with up to five slits. (a) Sketch of the fivefold slit and the detection configuration: one detector is fixed ($\delta_1=0$) while the other one is scanned ($\delta_2=\delta$). (b)-(e) Theoretically evaluated second- to fifth-order two-particle interference, Eq.~\eqref{eq:semi-general-IM} (normalized by the central peak), as a function of the optical phase $\delta$ for the detection technique indicated in (a). The coincident detection signal arises (f) in a double slit setup from four and (g) in a triple slit setup from nine mutually exclusive two-particle paths which can interfere.}
\end{figure*}

In case of many-particle phenomena, it is convenient to work within the second quantization framework, where the tedious symmetrization of wave functions is inherent in the operators. This will lead us to work with correlation functions.
The $M$th-order intensity correlation function is defined by~\cite{Glauber:1963}
\begin{eqnarray}\label{eq:GM}
G^{(M)}(\mathbf{r}_1,t_1,\ldots,\mathbf{r}_M,&&t_M) = \braket{\hat{a}^\dagger_1 \ldots \hat{a}^\dagger_M \hat{a}_M \ldots \hat{a}_1} \, ,
\end{eqnarray}
where $\hat{a}_i=\hat{a}(\mathbf{r}_i,t_i)$ is the bosonic annihilation operator at the position $\mathbf{r}_i$ of the $i$th detector and time of detection $t_i$. The function $G^{(M)}$ can be interpreted as the joint probability to detect $M$ photons at positions $\mathbf{r}_1,\ldots,\mathbf{r}_M$ and times $t_1,\ldots,t_M$, that is, a correlated detection at different detectors as shown in Fig.~\ref{fig:many-particle-interference}(a).
Since we work in the far field, contributions to the signal due to exotic looped paths through the slits~\cite{Yabuki:1986,Sawant:2014,Magana-Loaiza:2016} are negligible.
Throughout this study, we will concentrate on coincident events, $t_i=t$ and  omit the time (and position) dependence for simplicity.
We note that  the first-order correlation function $G^{(1)}_X$ is equal to the detection probability $P_X$ for any slit combination $X$, when evaluated for a single-particle wave function $\psi_X$. Hence, all  single-particle interference results can be analyzed in terms of the first-order correlation function $G^{(1)}$.

In order to examine many-particle interference, we assume, as in the case of single-particle interference, that the wave function of the $M$ particles is spatially coherent over the separation of the slits.
At first, we note that for a single slit $A$, we have $G^{(M)}_A= \prod_{i=1}^M G_A^{(1)}(\mathbf{r}_i) \equiv \prod_{i=1}^M G_{A,\text{cl.}}^{(1)}(\mathbf{r}_i) = G^{(M)}_{A,\text{cl.}}$, since there is no interference in single-slit experiments even when using quantum particles. 
The first configuration to exhibit (many-particle) interference is the double slit. To trigger a coincident event at $M$ detectors behind a double slit, there are in total $2^M$ distinct $M$-particle paths. Classically, all these events are mutually exclusive and the classical signal is simply given by the incoherent sum of the corresponding contributions,
\begin{eqnarray}
G^{(M)}_{AB,\text{cl.}} &=& G^{(M)}_A   \\
&&\hspace{1mm} + \sum_{j=1}^M G^{(M-1)}_{A}(\{\mathbf{r}_{i\neq j}\}_{i=1,\ldots,M}) G^{(1)}_{B}(\mathbf{r}_j) \notag \\
&&\hspace{1mm} + \ldots + G^{(M)}_{B} \, . \notag
\end{eqnarray}
Here, the first term represents the event, in which all detected particles came from $A$, the second expression includes all paths, in which one of the detected particles was from $B$ and $M-1$ from $A$, and so on.
The usual interference term, i.e., here the second-order $M$-particle interference, is defined in analogy to Eq.~\eqref{eq:I1_AB} as
\begin{eqnarray}\label{eq:IM_AB}
I^{(M)}_2  &\vcentcolon =& G^{(M)}_{AB} - G^{(M)}_{AB,\text{cl.}}  \, ,
\end{eqnarray}
where $G^{(M)}_{AB}$ is the quantum-mechanical result of Eq.~\eqref{eq:GM} including the interference between all occurring $M$-particle paths.

Following Sorkin, we can now generalize this second-order many-particle interference term to an arbitrary order of interference. 
In doing so, it becomes apparent that we can assign a great number of occurring terms (but not all) to either arising from classical terms or from interferences of lower order~\cite{Supplemental}. 
Consequently, we can define a hierarchy of higher-order many-particle interference terms containing the particular interference of $N$th order that was not present in \emph{any} lower-order many-particle interference. 
In this way, the $N$th-order $M$-particle interference can be written as~\cite{Supplemental}
\begin{eqnarray}\label{eq:semi-general-IM}
I^{(M)}_N &\vcentcolon =& G^{(M)}_{A_1,A_2,\ldots,A_N}  \\
&&+ \sum_{l=1}^{N-1} (-1)^l \hspace{-1mm} \sum_{\sigma \in \mathcal{C}^N_{N-l}} \hspace{-1mm}  G^{(M)}_{\underbrace{A_{\sigma_1},A_{\sigma_2},\ldots,A_{\sigma_{N-l}}}_{(N-l) \text{ slits}}} \notag \\
&&-\hspace{1mm} \mathcal{G}^{(M)}_{A_1,A_2,\ldots,A_N,\text{cl.}} \notag  \, .
\end{eqnarray}
Equation~\eqref{eq:semi-general-IM} is straightforward to comprehend by considering interference patterns from $N$-slit experiments:
The first term represents the quantum-mechanical signal from $N$ slits, see Eq.~\eqref{eq:GM}. 
The expression in the second line of Eq.~\eqref{eq:semi-general-IM} subtracts from $G^{(M)}_{A_1,A_2,\ldots,A_N}$ the contributions to the signal of $N$ slits from lower-order terms (including interference from all comprised $k$-slit configurations with $k<N$), where $\mathcal{C}_k^N$ stands for the set of all $k$-combinations of $\{1,2,\ldots ,N\}$. 
In the third line, we further subtract from $G^{(M)}_{A_1,A_2,\ldots,A_N}$ via $\mathcal{G}$  the classical $M$-particle paths arising from exactly $N$ different slits~\cite{Supplemental}.
Note that Eq.~\eqref{eq:semi-general-IM} is defined such that it is also valid for the single-particle case $M=1$, see Eq.~\eqref{eq:I1_AB}.
The formal resemblance of single- and many-particle interference hierarchies originates from a similar packing of the occurring terms. However, they differ strongly. The second-order two-particle interference $I^{(2)}_2$, for instance, is composed of up to $12$ single terms compared to only $2$ for the  second-order single-particle interference $I^{(1)}_2$~\cite{Supplemental}.

While the third-order {single-particle} interference term is the first to vanish in the family of {single-particle} interference terms, {many-particle} interference yields a much richer structure.
For example, for two-particle interference ($M=2$) after a double slit ($N=2$) 
there are $N^{M}=2^2=4$ different two-particle paths labeling the different options how two indistinguishable particles may reach two detectors, namely the detected particles are either (i) both from A, or (ii) both from B, or one from A and one from B, where (iii) detector 1 registered the particle from A and detector 2 registered the particle from B, or (iv) detector 1 registered the particle from B and detector 2 registered the particle from A, respectively. These paths are shown in Fig.~\ref{fig:many-particle-interference}(f).
The classical signal can thus be written as,
$G^{(2)}_{AB,\text{cl.}} = G^{(2)}_A  + G^{(2)}_B  + G^{(1)}_A(\mathbf{r}_1) G^{(1)}_B(\mathbf{r}_2) + G^{(1)}_A(\mathbf{r}_2) G^{(1)}_B(\mathbf{r}_1)$,
and the second-order {two-particle} interference is given by $I^{(2)}_2  = G^{(2)}_{AB} - G^{(2)}_{AB,\text{cl.}}$, which is nonzero~\cite{Supplemental}.
Considering the next order $I^{(2)}_3$, it turns out that two-particle interference in a triple-slit-experiment
can, unlike the single-particle analog, not be decomposed into a sum of lower-order two-particle interference patterns, even though all interfering terms shown in Fig.~\ref{fig:many-particle-interference}(g) do already appear in the appropriate lower-order ones \footnote{The two-particle events (i, ii, iii) of Fig.~\ref{fig:many-particle-interference}(g) do appear in single as well as double slit interference, while (iv, v), (vi, vii) and (viii, ix) appear in the double slits $AB$, $AC$ and $BC$, respectively.}. Third-order two-particle interference hence does not vanish, which holds also true for the fourth-order: $I^{(2)}_3\neq 0$ and $I^{(2)}_4\neq 0$~\cite{Supplemental}. The fifth-order two-particle interference, however, is now the first of the family members to vanish $I^{(2)}_5=0$, and thus all $I^{(2)}_N=0$ for $N\geq 5$. 

For this two-particle case, we have theoretically evaluated the two-particle interference terms of up to fifth-order for the detector configuration shown in Fig.~\ref{fig:many-particle-interference}(a). In this setup, $(M-1)$ detectors are located at fixed positions, while the $M$th detector scans the detection plane. This technique is commonly used to explore many-particle phenomena, such as superradiance~\cite{Oppel:2014} and subradiance~\cite{Bhatti:2018}. It is also used for quantum imaging~\cite{Schneider:2017}. Our findings for $M=2$ are shown in Fig.~\ref{fig:many-particle-interference}(b)-(e). Here, the respective interferences are plotted as a function of the optical phases $\delta_i(\mathbf{r}_i)=2\pi d \sin (\theta_i) /\lambda$ involving the slit distance $d$, the angle of detection $\theta_i$ of the $i$th detector and the wavelength $\lambda$. As predicted by our theory, we observe $I^{(2)}_5=0$, which is independent of the actual detector configuration~\cite{Supplemental}.

We stress that non-vanishing third- and fourth-order two-particle interference is in line with Born's rule, which in this context  predicts that quantum interference originates from pairs of two-particle paths, which is why fifth-order two-particle interference has to vanish.
Generally speaking, Born's law thus predicts that quantum interferences arise from pairs of $M$-particle paths, i.e., up to order $2M$. As a result,  by recursion all  interferences of order \mbox{$(2M+1)$} and larger are zero for $M$ particles.

Accordingly, we can now introduce a family of Sorkin parameters for $M$ particles,
\begin{eqnarray}\label{eq:Sorkin-family}
\kappa^{(M)}(\mathbf{r}_1,\ldots,\mathbf{r}_M):=\frac{I^{(M)}_{2M+1}(\mathbf{r}_1,\ldots,\mathbf{r}_M)}{G^{(M)}_{A_1,\ldots,A_{2M+1}}(0)} \, ,
\end{eqnarray}
where the first member ($M=1$) is the original Sorkin parameter of Eq.~\eqref{eq:Sorkin-parameter}. Born's law  predicts that all of these parameters vanish regardless of the detector positions.

\paragraph*{Sensitivity to deviations.}

We next derive that the many-particle Sorkin parameters [Eq.~\eqref{eq:Sorkin-family}] are more sensitive to deviations from Born's rule than the corresponding  single-particle parameter [Eq.~\eqref{eq:Sorkin-parameter}].
To this end, we assume that the probability of an event $A$ is given by $P_A=|\psi_A|^2 + \Delta_A$, where $\Delta_A$ represents a generic  deviation from Born's law  which we do not specify further for the sake of generality. 
A relation of the form, $P\propto |\psi|^{2+\epsilon}$, with $\epsilon \ll1 $, is for example of this type to first order in $\epsilon$. Since these deviations are expected to be very small, we assume that they are  of the same order of magnitude $\mathcal{O}(\Delta_X)\approx \Delta$, independent of $X$,  for different slit combinations $X$.
Such non-vanishing deviations from Born's rule will translate into nonzero Sorkin parameters of Eq.~\eqref{eq:Sorkin-family}. 
The third-order single-particle interference term in the numerator of the original Sorkin parameter [Eq.~\eqref{eq:Sorkin-parameter}] involves seven applications of the Born law. By conventional propagation of uncertainty~\cite{Barlow:1988}, the  deviations $\Delta$ lead to a nonzero single-particle Sorkin parameter of order of magnitude given by $\kappa^{(1)} \approx \sqrt{7}\Delta/G^{(1)}_{ABC}(0)$. 

\begin{figure}
	\centering \includegraphics[width=0.97\columnwidth]{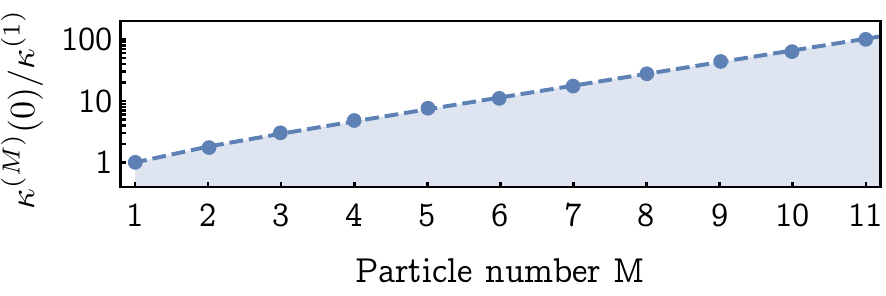}
	\caption{\label{fig:sensitivity} Exponential increase in sensitivity, $\kappa^{(M)}(0)/\kappa^{(1)}$, Eq.~\eqref{eq:sensitivity}, of the $M$-particle Sorkin parameter to deviations from Born's rule with respect to single-particle interference.}
	
\end{figure}

In the many-particle case, we evaluate the corresponding deviations in the generalized Sorkin parameters of Eq.~\eqref{eq:Sorkin-family}
for the commonly used detection configuration described above and outlined for $M=2$ in Fig.~\ref{fig:many-particle-interference}(a). We choose the fixed positions, $\mathbf{r}_i, \, (i = 1, \ldots , M-1)$, of the $(M-1)$ detectors such that the first detector is on the optical axis [see Fig.~\ref{fig:many-particle-interference}~(a)] and the optical phase difference accumulated by particles originating from the same slit but propagating to different detectors is always a multiple of $2\pi$. This can, for instance, be achieved with $\delta_i(\mathbf{r}_i)=(i-1) 2\pi$. Limiting ourselves to the central point of the $M$-particle Sorkin parameter, we can then relate the $M$th-order correlation function $G^{(M)}$ to the $M$th power of the central peak of the first-order correlation function $[G^{(1)}(0)]^M$~\cite{Supplemental}.
This enables us to transfer the deviations $\Delta$, discussed above for $G^{(1)}$, to $G^{(M)}$ and insert them into the definition of the \mbox{$M$-particle} Sorkin parameter Eq.~\eqref{eq:Sorkin-family}.
In this way, we obtain~\cite{Supplemental}
\begin{eqnarray}\label{eq:sensitivity}
\kappa^{(M)}(0) =  \frac{M \sqrt{C(M)} }{G^{(1)}_{A_1,\ldots,A_{2M+1}}(0)} \times \Delta + \mathcal{O}(\Delta^2)
\end{eqnarray}
for $M>1$. 
The  function $C(M)$ is related to the number of interfering $M$-particle paths in all possible $k$-slit combinations and reads~\cite{Supplemental}
\begin{eqnarray}
C(M) = \sum_{k=1}^{2M+1} \begin{pmatrix} 2M+1 \\ k \end{pmatrix} \left(\frac{k}{2M+1}\right)^{M-1}  \, .
\end{eqnarray}
Equation \eqref{eq:sensitivity} expresses the sensitivity to deviations from Born's rule for many-particle interference. 
The ratio to the sensitivity of the single-particle Sorkin parameter, $\kappa^{(M)}(0)/\kappa^{(1)}$, is independent of the original magnitude of the deviation $\Delta$ and thus constitutes a suitable measure to display the increase in sensitivity for many-particle interference with respect to single-particle interference. It is shown as a function of $M$ in Fig.~\ref{fig:sensitivity}. 
Remarkably, the increase in sensitivity depends exponentially on the number of particles $M$. This is linked to the exponential increase in the number of interfering paths. We observe an increase in sensitivity by a factor of about $2$ for the discussed two-particle example, by one order of magnitude for six-particle, and two orders of magnitude for eleven-particle interference. We note that the level of many-particle interference is not limited in principle. Eight-particle interference has, for instance, already been measured for superradiant emission~\cite{Oppel:2014}.

\paragraph*{Conclusion.}

We have shown that many-particle interference can be fruitfully exploited to test Born's rule. We have first established that all  interference of order $(2M+1)$ and larger vanish for $M$ particles. We have further introduced generalized Sorkin parameters applicable for many-particle interference which are predicted to be zero if Born's rule holds. The $M$-particle Sorkin parameters have the benefit of displaying an exponentially increased sensitivity to deviations from Born's rule with respect to their single-particle counterpart. 
We therefore expect them to stimulate new theoretical and experimental studies and more precise tests of quantum theory.


\vspace{0.3cm}
M.-O.P. gratefully acknowledges financial support by the Studienstiftung des deutschen Volkes. 
M.-O.P. and J.v.Z. gratefully acknowledge funding by the Erlangen Graduate School in Advanced Optical Technologies (SAOT) by the German Research Foundation (DFG) in the framework of the German excellence initiative.

\end{document}


\title{Supplemental Material: Many-particle interference to test Born's rule}

\author{Marc-Oliver Pleinert}
\affiliation{Institut f\"{u}r Optik, Information und Photonik, \\ Friedrich-Alexander-Universit\"{a}t Erlangen-N\"{u}rnberg (FAU), 91058 Erlangen, Germany}
\affiliation{Erlangen Graduate School in Advanced Optical Technologies (SAOT), Friedrich-Alexander-Universit\"{a}t Erlangen-N\"{u}rnberg (FAU), 91052 Erlangen, Germany}

\author{Joachim von Zanthier}
\affiliation{Institut f\"{u}r Optik, Information und Photonik, \\ Friedrich-Alexander-Universit\"{a}t Erlangen-N\"{u}rnberg (FAU), 91058 Erlangen, Germany}
\affiliation{Erlangen Graduate School in Advanced Optical Technologies (SAOT), Friedrich-Alexander-Universit\"{a}t Erlangen-N\"{u}rnberg (FAU), 91052 Erlangen, Germany}
\author{Eric Lutz}
\affiliation{Institute for Theoretical Physics I, University of Stuttgart, D-70550 Stuttgart, Germany}

\maketitle

\section{Classical contributions}\label{sec:classical-contr}

In a general setup with $M$ detectors behind $N$ slits, there are $N^M$ different $M$-particle paths to trigger an $M$-fold event. 
As classical particles are stochastically independent and display no interference, the classical signal of an $M$-fold event behind $N$ slits is given by the incoherent sum of these contributions,
\begin{eqnarray}\label{eq:GM-classical}
 G^{(M)}_{A_1,\ldots,A_N,\text{cl.}}  
= \sum_{\sigma \in \mathcal{P}_N^M} G_{A_{\sigma_1}}^{(1)}(\mathbf{r}_1) \ldots G_{A_{\sigma_N}}^{(1)}(\mathbf{r}_M)  \, , 
\end{eqnarray}
where the sum runs over all $M$-permutations with repetition of the set $\{1,\ldots,N\}$ ($=:\mathcal{P}_N^M$). 
The first term ($\sigma=\{1,1,\ldots\}$), for instance, represents the event, in which all detected particles came from slit $A_1$. 
To clearly differentiate between classical and quantum contributions later on, we group all classical $M$-particle paths that are of $N$th-order, i.e., they originate from exactly $N$ different slits, in the expression $\mathcal{G}^{(M)}_{A_1,\ldots,A_N,\text{cl.}}$, which can be defined inductively via
\begin{subequations}
	\begin{align}
	N=1&: & \mathcal{G}^{(M)}_{A_1,\text{cl.}} &\vcentcolon= G^{(M)}_{A_1,\text{cl.}} = G^{(M)}_{A_1}  \, , \\
	N\rightarrow N+1&: & \mathcal{G}^{(M)}_{A_1,A_2,\ldots,A_{N+1},\text{cl.}} &\vcentcolon=G^{(M)}_{A_1,A_2,\ldots,A_{N+1},\text{cl.}} 
	- \sum_{l=1}^{N} \sum_{\sigma \in \mathcal{C}^{N+1}_{l}}
	\mathcal{G}^{(M)}_{A_{\sigma_1},A_{\sigma_2},\ldots,A_{\sigma_l},\text{cl.}}  \, .
	\end{align}
\end{subequations}
In the induction step, the classical signal of order $(N+1)$ is calculated and afterwards all classical contributions of lower order (from less than $(N+1)$ slits) are gradually subtracted. Here, $\mathcal{C}^{N+1}_{l}$ stands for the set of all $l$-combinations of the set $\{1,2,\ldots ,N+1\}$ and accounts for all possible slit-configurations of exactly $l$ slits (out of $N+1$ slits).
We note a few things here:
\begin{itemize}
	\item The difference between $\mathcal{G}^{(M)}_{A_1,\ldots,A_{N},\text{cl.}}$ and $G^{(M)}_{A_1,\ldots,A_{N},\text{cl.}} $ is that the latter includes all classical paths of any order to the signal, while the former includes classical paths of $N$th-order only, i.e., from exactly $N$ different slits.
	\item $G^{(M)}_{A,cl.} = G^{(M)}_{A} $, since there is no interference in single-slit experiments even when using quantum particles. 
	\item $\mathcal{G}^{(M)}_{A_1,\ldots,A_{N},\text{cl.}}=0$ for $N>M$, i.e., when the number of slits $N$ exceeds the correlation order $M$. In this case $G^{(M)}_{A_1,\ldots,A_{N},\text{cl.}}$ can be written as the sum of lower-order classical terms, just as the classical single-particle double-slit term ($N=2>1=M$) can be decomposed into the sum of the two single slits and hence $\mathcal{G}_{AB}^{(1)}=0$.
\end{itemize}

\section{Two-particle interference patterns and hierarchy}

\begin{figure}
	\centering \includegraphics[width=\textwidth]{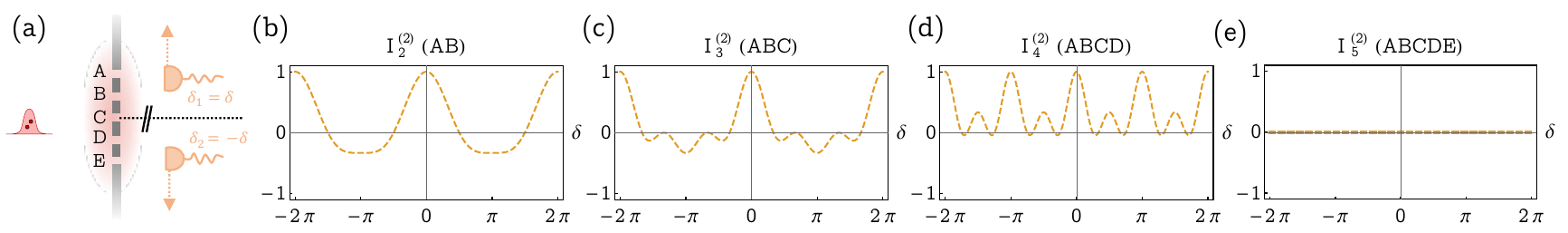}
	\caption{\label{fig:2nd-config} Two-particle interference in a second detector configuration. (a) Sketch of the fivefold slit and the further detection configuration, where the detectors are scanned in opposite directions ($\delta_1=\delta, \delta_2=-\delta$). (b)-(e) Theoretically evaluated second- to fifth-order two-particle interference (normalized by the central peak), as a function of the optical phase $\delta$ for the detection technique indicated in (a).}
\end{figure}

In this section, we derive the two-particle interference terms for double and triple slit in order to explicitly state third-order two-particle interference $I^{(2)}_{3} \neq 0$. We will use $\hat{A}$, $\hat{B}$, $\hat{C}$ to indicate the annihilation operator of photons from slit $A$, $B$, $C$, respectively, and write, like in the main text, $\hat{X}_i=\hat{X}(\mathbf{r}_i)$. Note, however, that actually $\hat{A}=\hat{B}=\hat{C}$, since we consider particles being spatially coherent over the slits such that all occurring quantum paths do interfere. We nevertheless distinguish the operators in order to discriminate between different two-particle paths and identify classically expected terms as well as lower-order interference terms. 
For a double slit, we get
\begin{eqnarray}\label{G2-AB-suppl}
G^{(2)}_{AB} &=\hphantom{\vcentcolon}& \braket{\hat{a}^\dagger_{AB}(\mathbf{r}_1)\hat{a}^\dagger_{AB}(\mathbf{r}_2)\hat{a}_{AB}(\mathbf{r}_2)\hat{a}_{AB}(\mathbf{r}_1)} \\
 &=\hphantom{\vcentcolon}& \langle (\hat{A}_1^\dagger  + \hat{B}_1^\dagger )(\hat{A}_2^\dagger  + \hat{B}_2^\dagger ) (\hat{A}_2 + \hat{B}_2)(\hat{A}_1 + \hat{B}_1) \rangle \notag \\
 &=\hphantom{\vcentcolon}& \langle \textcolor{blue}{\hat{A}_1^\dagger \hat{A}_2^\dagger \hat{A}_2\hat{A}_1} \rangle + \langle \hat{A}_1^\dagger \hat{A}_2^\dagger \hat{A}_2\hat{B}_1 \rangle + \langle \hat{A}_1^\dagger \hat{A}_2^\dagger \hat{B}_2\hat{A}_1 \rangle + \langle \hat{A}_1^\dagger \hat{A}_2^\dagger \hat{B}_2\hat{B}_1 \rangle \notag + \langle \hat{A}_1^\dagger \hat{B}_2^\dagger \hat{A}_2\hat{A}_1 \rangle + \langle \hat{A}_1^\dagger \hat{B}_2^\dagger \hat{A}_2\hat{B}_1 \rangle \\
 &&+ \langle \textcolor{red}{\hat{A}_1^\dagger \hat{B}_2^\dagger \hat{B}_2\hat{A}_1} \rangle + \langle \hat{A}_1^\dagger \hat{B}_2^\dagger \hat{B}_2\hat{B}_1 \rangle + \langle \hat{B}_1^\dagger \hat{A}_2^\dagger \hat{A}_2\hat{A}_1 \rangle + \langle \textcolor{red}{\hat{B}_1^\dagger \hat{A}_2^\dagger \hat{A}_2\hat{B}_1} \rangle + \langle \hat{B}_1^\dagger \hat{A}_2^\dagger \hat{B}_2\hat{A}_1 \rangle + \langle \hat{B}_1^\dagger \hat{A}_2^\dagger \hat{B}_2\hat{B}_1 \rangle \notag \\
 &&+ \langle \hat{B}_1^\dagger \hat{B}_2^\dagger \hat{A}_2\hat{A}_1 \rangle + \langle \hat{B}_1^\dagger \hat{B}_2^\dagger \hat{A}_2\hat{B}_1 \rangle + \langle \hat{B}_1^\dagger \hat{B}_2^\dagger \hat{B}_2\hat{A}_1 \rangle + \langle \textcolor{blue}{\hat{B}_1^\dagger B_2^\dagger B_2B_1} \rangle \notag \\
 &=\vcentcolon& \textcolor{blue}{G^{(2)}_A} + \textcolor{blue}{G^{(2)}_B} + \textcolor{red}{G^{(1)}_A(\mathbf{r}_1) G^{(1)}_B(\mathbf{r}_2) + G^{(1)}_A(\mathbf{r}_2) G^{(1)}_B(\mathbf{r}_1)} + I^{(2)}_{AB} \, , \notag
\end{eqnarray}
where we marked the classically expected contributions in red and blue (see main text) and eventually defined the remaining terms as second-order two-particle interference $I^{(2)}_{AB} \equiv I^{(2)}_2$. Note that throughout the Supplemental Material we explicitly state the involved slits for convenience as $A, B, \ldots$, where the order of interference is included inherently already in the number of slits.
In the same manner, we get for a triple slit 
\begin{eqnarray}\label{G2-ABC-suppl}
G^{(2)}_{ABC} &=\hphantom{\vcentcolon}& \braket{\hat{a}^\dagger_{ABC}(\mathbf{r}_1)\hat{a}^\dagger_{ABC}(\mathbf{r}_2)\hat{a}_{ABC}(\mathbf{r}_2)\hat{a}_{ABC}(\mathbf{r}_1)}  \\
 &=\hphantom{\vcentcolon}& \langle (\hat{A}_1^\dagger + \hat{B}_1^\dagger + \hat{C}_1^\dagger )(\hat{A}_2^\dagger  + \hat{B}_2^\dagger  + \hat{C}_2^\dagger ) (\hat{A}_2 + \hat{B}_2 + \hat{C}_2)(\hat{A}_1 + \hat{B}_1 + \hat{C}_1) \rangle \notag \\
 &=\hphantom{\vcentcolon}& \langle \textcolor{blue}{\hat{A}_1^\dagger   \hat{A}_2^\dagger   \hat{A}_2  \hat{A}_1} \rangle + 
 \langle \textcolor{violet}{\hat{A}_1^\dagger   \hat{A}_2^\dagger   \hat{A}_2  \hat{B}_1} \rangle + 
 \langle \textcolor{violet}{\hat{A}_1^\dagger   \hat{A}_2^\dagger   \hat{A}_2  \hat{C}_1} \rangle + 
 \langle \textcolor{violet}{\hat{A}_1^\dagger   \hat{A}_2^\dagger   \hat{B}_2  \hat{A}_1} \rangle \notag + 
 \langle \textcolor{violet}{\hat{A}_1^\dagger   \hat{A}_2^\dagger   \hat{B}_2  \hat{B}_1} \rangle + 
 \langle \hat{A}_1^\dagger   \hat{A}_2^\dagger   \hat{B}_2  \hat{C}_1 \rangle \\ &&+ 
 \langle \textcolor{violet}{\hat{A}_1^\dagger   \hat{A}_2^\dagger   \hat{C}_2  \hat{A}_1} \rangle + 
 \langle \hat{A}_1^\dagger   \hat{A}_2^\dagger   \hat{C}_2  \hat{B}_1 \rangle \notag + 
 \langle \textcolor{violet}{\hat{A}_1^\dagger   \hat{A}_2^\dagger   \hat{C}_2  \hat{C}_1} \rangle + 
 \langle \textcolor{violet}{\hat{A}_1^\dagger   \hat{B}_2^\dagger   \hat{A}_2  \hat{A}_1} \rangle + 
 \langle \textcolor{violet}{\hat{A}_1^\dagger   \hat{B}_2^\dagger   \hat{A}_2  \hat{B}_1} \rangle + 
 \langle \hat{A}_1^\dagger   \hat{B}_2^\dagger   \hat{A}_2  \hat{C}_1 \rangle \notag \\ &&+ 
 \langle \textcolor{red}{\hat{A}_1^\dagger   \hat{B}_2^\dagger   \hat{B}_2  \hat{A}_1} \rangle + 
 \langle \textcolor{violet}{\hat{A}_1^\dagger   \hat{B}_2^\dagger   \hat{B}_2  \hat{B}_1} \rangle + 
 \langle \hat{A}_1^\dagger   \hat{B}_2^\dagger   \hat{B}_2  \hat{C}_1 \rangle + 
 \langle \hat{A}_1^\dagger   \hat{B}_2^\dagger   \hat{C}_2  \hat{A}_1 \rangle \notag+ 
 \langle \hat{A}_1^\dagger   \hat{B}_2^\dagger   \hat{C}_2  \hat{B}_1 \rangle + 
 \langle \hat{A}_1^\dagger   \hat{B}_2^\dagger   \hat{C}_2  \hat{C}_1 \rangle \\ &&+ 
 \langle \textcolor{violet}{\hat{A}_1^\dagger   \hat{C}_2^\dagger   \hat{A}_2  \hat{A}_1} \rangle + 
 \langle \hat{A}_1^\dagger   \hat{C}_2^\dagger   \hat{A}_2  \hat{B}_1 \rangle \notag + 
 \langle \textcolor{violet}{\hat{A}_1^\dagger   \hat{C}_2^\dagger   \hat{A}_2  \hat{C}_1} \rangle + 
 \langle \hat{A}_1^\dagger   \hat{C}_2^\dagger   \hat{B}_2  \hat{A}_1 \rangle + 
 \langle \hat{A}_1^\dagger   \hat{C}_2^\dagger   \hat{B}_2  \hat{B}_1 \rangle + 
 \langle \hat{A}_1^\dagger   \hat{C}_2^\dagger   \hat{B}_2  \hat{C}_1 \rangle \notag \\ &&+ 
 \langle \textcolor{red}{\hat{A}_1^\dagger   \hat{C}_2^\dagger   \hat{C}_2  \hat{A}_1} \rangle + 
 \langle \hat{A}_1^\dagger   \hat{C}_2^\dagger   \hat{C}_2  \hat{B}_1 \rangle + 
 \langle \textcolor{violet}{\hat{A}_1^\dagger   \hat{C}_2^\dagger   \hat{C}_2  \hat{C}_1} \rangle + 
 \langle \textcolor{violet}{\hat{B}_1^\dagger   \hat{A}_2^\dagger   \hat{A}_2  \hat{A}_1} \rangle \notag + 
 \langle \textcolor{red}{\hat{B}_1^\dagger   \hat{A}_2^\dagger   \hat{A}_2  \hat{B}_1} \rangle + 
 \langle \hat{B}_1^\dagger   \hat{A}_2^\dagger   \hat{A}_2  \hat{C}_1 \rangle \\ &&+ 
 \langle \textcolor{violet}{\hat{B}_1^\dagger   \hat{A}_2^\dagger   \hat{B}_2  \hat{A}_1} \rangle + 
 \langle \textcolor{violet}{\hat{B}_1^\dagger   \hat{A}_2^\dagger   \hat{B}_2  \hat{B}_1} \rangle \notag + 
 \langle \hat{B}_1^\dagger   \hat{A}_2^\dagger   \hat{B}_2  \hat{C}_1 \rangle + 
 \langle \hat{B}_1^\dagger   \hat{A}_2^\dagger   \hat{C}_2  \hat{A}_1 \rangle + 
 \langle \hat{B}_1^\dagger   \hat{A}_2^\dagger   \hat{C}_2  \hat{B}_1 \rangle + 
 \langle \hat{B}_1^\dagger   \hat{A}_2^\dagger   \hat{C}_2  \hat{C}_1 \rangle \notag \\ &&+ 
 \langle \textcolor{violet}{\hat{B}_1^\dagger   \hat{B}_2^\dagger   \hat{A}_2  \hat{A}_1} \rangle + 
 \langle \textcolor{violet}{\hat{B}_1^\dagger   \hat{B}_2^\dagger   \hat{A}_2  \hat{B}_1} \rangle + 
 \langle \hat{B}_1^\dagger   \hat{B}_2^\dagger   \hat{A}_2  \hat{C}_1 \rangle + 
 \langle \textcolor{violet}{\hat{B}_1^\dagger   \hat{B}_2^\dagger   \hat{B}_2  \hat{A}_1} \rangle \notag+ 
 \langle \textcolor{blue}{\hat{B}_1^\dagger   \hat{B}_2^\dagger   \hat{B}_2  \hat{B}_1} \rangle + 
 \langle \textcolor{violet}{\hat{B}_1^\dagger   \hat{B}_2^\dagger   \hat{B}_2  \hat{C}_1} \rangle \\ &&+ 
 \langle \hat{B}_1^\dagger   \hat{B}_2^\dagger   \hat{C}_2  \hat{A}_1 \rangle + 
 \langle \textcolor{violet}{\hat{B}_1^\dagger   \hat{B}_2^\dagger   \hat{C}_2  \hat{B}_1} \rangle \notag + 
 \langle \textcolor{violet}{\hat{B}_1^\dagger   \hat{B}_2^\dagger   \hat{C}_2  \hat{C}_1} \rangle + 
 \langle \hat{B}_1^\dagger   \hat{C}_2^\dagger   \hat{A}_2  \hat{A}_1 \rangle + 
 \langle \hat{B}_1^\dagger   \hat{C}_2^\dagger   \hat{A}_2  \hat{B}_1 \rangle + 
 \langle \hat{B}_1^\dagger   \hat{C}_2^\dagger   \hat{A}_2  \hat{C}_1 \rangle \notag \\ &&+ 
 \langle \hat{B}_1^\dagger   \hat{C}_2^\dagger   \hat{B}_2  \hat{A}_1 \rangle + 
 \langle \textcolor{violet}{\hat{B}_1^\dagger   \hat{C}_2^\dagger   \hat{B}_2  \hat{B}_1} \rangle + 
 \langle \textcolor{violet}{\hat{B}_1^\dagger   \hat{C}_2^\dagger   \hat{B}_2  \hat{C}_1} \rangle + 
 \langle \hat{B}_1^\dagger   \hat{C}_2^\dagger   \hat{C}_2  \hat{A}_1 \rangle \notag + 
 \langle \textcolor{red}{\hat{B}_1^\dagger   \hat{C}_2^\dagger   \hat{C}_2  \hat{B}_1} \rangle + 
 \langle \textcolor{violet}{\hat{B}_1^\dagger   \hat{C}_2^\dagger   \hat{C}_2  \hat{C}_1} \rangle \\ &&+ 
 \langle \textcolor{violet}{\hat{C}_1^\dagger   \hat{A}_2^\dagger   \hat{A}_2  \hat{A}_1} \rangle + 
 \langle \hat{C}_1^\dagger   \hat{A}_2^\dagger   \hat{A}_2  \hat{B}_1 \rangle \notag + 
 \langle \textcolor{red}{\hat{C}_1^\dagger   \hat{A}_2^\dagger   \hat{A}_2  \hat{C}_1} \rangle + 
 \langle \hat{C}_1^\dagger   \hat{A}_2^\dagger   \hat{B}_2  \hat{A}_1 \rangle + 
 \langle \hat{C}_1^\dagger   \hat{A}_2^\dagger   \hat{B}_2  \hat{B}_1 \rangle + 
 \langle \hat{C}_1^\dagger   \hat{A}_2^\dagger   \hat{B}_2  \hat{C}_1 \rangle \notag \\ &&+ 
 \langle \textcolor{violet}{\hat{C}_1^\dagger   \hat{A}_2^\dagger   \hat{C}_2  \hat{A}_1} \rangle + 
 \langle \hat{C}_1^\dagger   \hat{A}_2^\dagger   \hat{C}_2  \hat{B}_1 \rangle + 
 \langle \textcolor{violet}{\hat{C}_1^\dagger   \hat{A}_2^\dagger   \hat{C}_2  \hat{C}_1} \rangle + 
 \langle \hat{C}_1^\dagger   \hat{B}_2^\dagger   \hat{A}_2  \hat{A}_1 \rangle \notag + 
 \langle \hat{C}_1^\dagger   \hat{B}_2^\dagger   \hat{A}_2  \hat{B}_1 \rangle + 
 \langle \hat{C}_1^\dagger   \hat{B}_2^\dagger   \hat{A}_2  \hat{C}_1 \rangle \\ &&+ 
 \langle \hat{C}_1^\dagger   \hat{B}_2^\dagger   \hat{B}_2  \hat{A}_1 \rangle + 
 \langle \textcolor{violet}{\hat{C}_1^\dagger   \hat{B}_2^\dagger   \hat{B}_2  \hat{B}_1} \rangle \notag + 
 \langle \textcolor{red}{\hat{C}_1^\dagger   \hat{B}_2^\dagger   \hat{B}_2  \hat{C}_1} \rangle + 
 \langle \hat{C}_1^\dagger   \hat{B}_2^\dagger   \hat{C}_2  \hat{A}_1 \rangle + 
 \langle \textcolor{violet}{\hat{C}_1^\dagger   \hat{B}_2^\dagger   \hat{C}_2  \hat{B}_1} \rangle + 
 \langle \textcolor{violet}{\hat{C}_1^\dagger   \hat{B}_2^\dagger   \hat{C}_2  \hat{C}_1} \rangle \notag \\ &&+ 
 \langle \textcolor{violet}{\hat{C}_1^\dagger   \hat{C}_2^\dagger   \hat{A}_2  \hat{A}_1} \rangle + 
 \langle \hat{C}_1^\dagger   \hat{C}_2^\dagger   \hat{A}_2  \hat{B}_1 \rangle + 
 \langle \textcolor{violet}{\hat{C}_1^\dagger   \hat{C}_2^\dagger   \hat{A}_2  \hat{C}_1} \rangle + 
 \langle \hat{C}_1^\dagger   \hat{C}_2^\dagger   \hat{B}_2  \hat{A}_1 \rangle \notag + 
 \langle \textcolor{violet}{\hat{C}_1^\dagger   \hat{C}_2^\dagger   \hat{B}_2  \hat{B}_1} \rangle + 
 \langle \textcolor{violet}{\hat{C}_1^\dagger   \hat{C}_2^\dagger   \hat{B}_2  \hat{C}_1} \rangle \\ &&+ 
 \langle \textcolor{violet}{\hat{C}_1^\dagger   \hat{C}_2^\dagger   \hat{C}_2  \hat{A}_1} \rangle + 
 \langle \textcolor{violet}{\hat{C}_1^\dagger   \hat{C}_2^\dagger   \hat{C}_2  \hat{B}_1} \rangle \notag + 
 \langle \textcolor{blue}{\hat{C}_1^\dagger   \hat{C}_2^\dagger   \hat{C}_2  \hat{C}_1} \rangle \notag \\
 &=\vcentcolon& \textcolor{blue}{G^{(2)}_A} + \textcolor{blue}{G^{(2)}_B} + \textcolor{blue}{G^{(2)}_C} + \textcolor{red}{G^{(1)}_A(\mathbf{r}_1) G^{(1)}_B(\mathbf{r}_2) + G^{(1)}_A(\mathbf{r}_2) G^{(1)}_B(\mathbf{r}_1)} + \textcolor{red}{G^{(1)}_A(\mathbf{r}_1) G^{(1)}_C(\mathbf{r}_2) + G^{(1)}_A(\mathbf{r}_2) G^{(1)}_C(\mathbf{r}_1)} \notag \\
 &&+ \textcolor{red}{G^{(1)}_B(\mathbf{r}_1) G^{(1)}_C(\mathbf{r}_2) + G^{(1)}_B(\mathbf{r}_2) G^{(1)}_C(\mathbf{r}_1)} + \textcolor{violet}{I^{(2)}_{AB}} + \textcolor{violet}{I^{(2)}_{AC}} + \textcolor{violet}{I^{(2)}_{BC}} + I^{(2)}_{ABC} \notag \\
 &=\hphantom{\vcentcolon}& G^{(2)}_{AB} + G^{(2)}_{AC} + G^{(2)}_{BC} - G^{(2)}_A - G^{(2)}_B - G^{(2)}_C + I^{(2)}_{ABC} \, , \notag
\end{eqnarray}
where we marked again all previously known lower-order terms, i.e., the classically expected contributions (in red and blue) and lower-order interference (in violet). Observe that here $\mathcal{G}^{(2)}_{ABC} = 0$ (since $M=2 < 3=N$) so that we do no have to identify emerging classical contributions. Eventually, we define the third-order two-particle interference via
\begin{eqnarray}\label{eq:I2_ABC_appendix}
I^{(2)}_{ABC} &\vcentcolon=& G^{(2)}_{ABC} - G^{(2)}_{AB} - G^{(2)}_{AC} - G^{(2)}_{BC} + G^{(2)}_A + G^{(2)}_B + G^{(2)}_C \, .
\end{eqnarray}
Hence, unlike zero single-particle third-order interference $I^{(1)}_{ABC}=0$ (see main text), we get a nonzero two-particle third-order interference $I^{(2)}_{ABC}\neq 0$. 
Defining two-particle interference of higher order is now straightforward (see also the next section). At last, we state the fifth-order two-particle interference term, as it will be the first in this hierarchy to become zero (see the main text):
\begin{eqnarray}\label{eq:I2_ABCDE_appendix}
I^{(2)}_{5}\equiv I^{(2)}_{ABCDE} &=& G^{(2)}_{ABCDE} - G^{(2)}_{ABCD} - G^{(2)}_{ABCE} - G^{(2)}_{ABDE} - G^{(2)}_{ACDE} - G^{(2)}_{BCDE} \\ 
&& + G^{(2)}_{ABC} + G^{(2)}_{ABD} + G^{(2)}_{ABE} + G^{(2)}_{ACD} + G^{(2)}_{ACE} + G^{(2)}_{ADE} + G^{(2)}_{BCD} + G^{(2)}_{BCE} + G^{(2)}_{BDE} + G^{(2)}_{CDE} \notag \\
&& - G^{(2)}_{AB} - G^{(2)}_{AC} - G^{(2)}_{AD} - G^{(2)}_{AE} - G^{(2)}_{BC} - G^{(2)}_{BD} - G^{(2)}_{BE} - G^{(2)}_{CD} - G^{(2)}_{CE} - G^{(2)}_{DE} \notag \\
&&+ G^{(2)}_{A} + G^{(2)}_{B} + G^{(2)}_{C} + G^{(2)}_{D} + G^{(2)}_{E} = 0 \, .\notag
\end{eqnarray}
$I^{(2)}_{5} = 0$ holds independent of the actual detector configuration, since the latter has not been involved in the derivation. This can be seen in Fig.~\ref{fig:2nd-config}, where we have theoretically evaluated the two-particle interference terms of up to fifth-order for a further detector configuration where detectors are scanned in opposite directions ($\delta_1=\delta, \delta_2=-\delta$), which has been used in Ref.~\cite{Thiel:2007} to overcome the classical resolution limit.

For the derived interference patterns, one has to track all possible interfering quantum paths, which lead to a coincident detection event, and their relative phase. In the general case of $M$-particle interference in $N$ slit-experiments, this can be done in the first quantization scheme by propagating the $M$-particle wave function, which is first coherently distributed over $N$ slits and afterwards detected coincidentally at the $M$ detectors. In the second quantization scheme, on the other hand, the relative phases are included in the operators. For example, $\hat{X}_i = \hat{X}(\mathbf{r}_i)$ from above, which is the annihilation operator of a photon traversing slit $X$ and being registered at the $i$th detector, inherently includes the optical phase $e^{ikr_{Xi}}$, where $k$ is the wave vector and $r_{Xi}$ is the distance between slit $X$ and the $i$th detector. We can thus exploit expressions like those in Eqs.~\eqref{G2-AB-suppl} and~\eqref{G2-ABC-suppl} to compute the interference terms by using $\hat{X}_i \propto e^{ikr_{Xi}} \hat{X}$.

\section{Interference hierarchy}

To derive the interference hierarchy, we start with the second-order, which is defined - analog to Sorkin - as the difference of the quantum and the classical result (as in the main text). From a given interference of order $(N-1)$, the next-higher interference of order $N$ can be obtained by calculating the quantum-mechanical result of slit experiments of this order, $G^{(M)}_{A_1,A_2,\ldots,A_N}$, and afterwards identifying and subtracting lower-order interference terms and classical terms. The climbing of the ladder of interference terms is, in principle, straightforward, the identification of lower-order terms at each level, however, can be tedious.

In the main text, the following compact expression of the interference hierarchy was given
\begin{eqnarray}\label{eq:main-text-expr}
I^{(M)}_N &=& G^{(M)}_{A_1,A_2,\ldots,A_N}  + \sum_{l=1}^{N-1} (-1)^l \hspace{-1mm} \sum_{\sigma \in \mathcal{C}^N_{N-l}}  G^{(M)}_{\underbrace{A_{\sigma_1},A_{\sigma_2},\ldots,A_{\sigma_{N-l}}}_{(N-l) \text{ slits}}} - \mathcal{G}^{(M)}_{A_1,A_2,\ldots,A_N,\text{cl.}}  \, ,
\end{eqnarray}
which displays the recipe from above. If we write out the sum,
\begin{eqnarray}
I^{(M)}_N &=& G^{(M)}_{A_1,A_2,\ldots,A_N}   \notag + (-1)^1 \sum_{\sigma \in \mathcal{C}^N_{N-1}}  G^{(M)}_{\underbrace{A_{\sigma_1},A_{\sigma_2},\ldots,A_{\sigma_{N-1}}}_{(N-1) \text{ slits}}} \notag + (-1)^2 \sum_{\sigma \in \mathcal{C}^N_{N-2}} G^{(M)}_{\underbrace{A_{\sigma_1},A_{\sigma_2},\ldots,A_{\sigma_{N-2}}}_{(N-2) \text{ slits}}}    \notag \\
&& + \ldots + (-1)^{N-1} \sum_n G^{(M)}_{A_n} \hspace{2mm} - \hspace{1mm} \mathcal{G}^{(M)}_{A_1,A_2,\ldots,A_N,\text{cl.}}  \notag  \, ,
\end{eqnarray}
one can clearly identify the different contributions: The first term on the right hand side contains all possible $N^{2M}$ terms (interferences between $N^M$ quantum paths) up to the $N$th order (also including the classical paths). The second expression subtracts terms up to $(N-1)$th order. In doing so, we subtract terms of \emph{up to} $(N-1)$th order, hence also of order $(N-2)$. In the calculation, it turns out that we actually subtract the latter contributions twice (loosely speaking, this is due to $G^{(M)}_{A}$ being present in $G^{(M)}_{AB}$ and $G^{(M)}_{AC}$, for instance). This is why we have to add them in the next step again. As a result we have to alternate between subtracting and adding lower-order terms (hence the factor $(-1)^l$ in the above equation). Equations~\eqref{eq:I2_ABC_appendix} and~\eqref{eq:I2_ABCDE_appendix} display this alternation for the third and fifth-order in the case of two-particle interference. These alternating terms account for lower-order interference and classical terms [up to order $(N-1)$]. Additionally arising classical terms of order $N$ are subtracted via the last term, $ \mathcal{G}^{(M)}_{A_1,A_2,\ldots,A_N,\text{cl.}}$, which is given in section~\ref{sec:classical-contr}.

\section{Derivation of the sensitivity to deviations}

In the final section, we derive the scaling of the many-particle Sorkin parameter to deviations from Born's rule. First we note that instead of using the actual position of the detectors, $\mathbf{r}_i$, we can also write the correlation functions in terms of the optical phases $\delta_i(\mathbf{r}_i)$, i.e., $G^{(1)}(\mathbf{r}_1,\ldots,\mathbf{r}_M) \equiv G^{(1)}(\delta_1,\ldots,\delta_M) $.
In the specific detector configuration, $\delta_i = (i-1)2\pi$ of the fixed $(M-1)$ detectors as stated in the main text, the central point of the $M$th-order correlation function reduces to the $M$th power of the central point of the first-order correlation function,
since all detectors are effectively located on the optical axis (due to the $2\pi$-symmetry). This allows us to directly transfer the deviations $\Delta$ (defined for $G^{(1)}$) to $M$th-order correlation functions $G^{(M)}$ and results in
\begin{eqnarray}\label{eq:GM-deviations}
G^{(M)}_{X}=\left(G^{(1)}_X\right)^M\rightarrow \left(G^{(1)}_X + \Delta\right)^M \approx G^{(M)}_{X} +M G^{(M-1)}_X\Delta + \mathcal{O}(\Delta^2) \, ,
\end{eqnarray}
where all terms are evaluated at $\delta_i = 0 \, (\mathbf{r}_i=0) \, \forall i$. In the following, we neglect terms $\mathcal{O}(\Delta^2)$ and higher. 

Let us first investigate the two-particle Sorkin parameter in detail to see the overall procedure, while we afterwards pass over to general $M$-particle interference. Inserting the above deviations into the central point of the two-particle Sorkin parameter leads to
\begin{eqnarray}\label{eq:kappa2}
\kappa^{(2)}(0) &=& 2   \Big[G^{(1)}_{ABCDE}\Delta_{ABCDE} \\
&& - G^{(1)}_{ABCD}\Delta_{ABCD} - G^{(1)}_{ABCE}\Delta_{ABCE} - G^{(1)}_{ABDE}\Delta_{ABDE}- G^{(1)}_{ACDE}\Delta_{ACDE} - G^{(1)}_{BCDE}\Delta_{BCDE} \notag \\ 
&& + G^{(1)}_{ABC}\Delta_{ABC} + G^{(1)}_{ABD}\Delta_{ABD} + G^{(1)}_{ABE}\Delta_{ABE} + G^{(1)}_{ACD}\Delta_{ACD} + G^{(1)}_{ACE}\Delta_{ACE} \notag \\
&&+ G^{(1)}_{ADE}\Delta_{ADE} + G^{(1)}_{BCD}\Delta_{BCD} + G^{(1)}_{BCE}\Delta_{BCE} + G^{(1)}_{BDE}\Delta_{BDE} + G^{(1)}_{CDE}\Delta_{CDE} \notag \\
&& - G^{(1)}_{AB}\Delta_{AB} - G^{(1)}_{AC}\Delta_{AC} - G^{(1)}_{AD}\Delta_{AD} - G^{(1)}_{AE}\Delta_{AE} - G^{(1)}_{BC}\Delta_{BC} \notag \\
&&- G^{(1)}_{BD}\Delta_{BD} - G^{(1)}_{BE}\Delta_{BE} - G^{(1)}_{CD}\Delta_{CD} - G^{(1)}_{CE}\Delta_{CE} - G^{(1)}_{DE}\Delta_{DE} \notag \\
&&+ G^{(1)}_{A}\Delta_{A} + G^{(1)}_{B}\Delta_{B} + G^{(1)}_{C}\Delta_{C}+ G^{(1)}_{D}\Delta_{D} + G^{(1)}_{E}\Delta_{E} \Big] \Big/ \left[G^{(1)}_{ABCDE}\cdot G^{(1)}_{ABCDE}\right]  \notag \\
&\approx& 2\frac{G^{(1)}_{ABCDE}\Delta_{ABCDE} - 5 G^{(1)}_{ABCD}\Delta_{ABCD} + 10 G^{(1)}_{ABC}\Delta_{ABC} - 10 G^{(1)}_{AB}\Delta_{AB} + 5 G^{(1)}_{A}\Delta_{A}}{G^{(1)}_{ABCDE}\cdot G^{(1)}_{ABCDE}} \, . \notag 
\end{eqnarray}
In a first step, we already neglected all $G^{(2)}$'s in the numerator (as they represent the part coming from Born's rule and thus cancel). In a second step, we eased the notation by assuming that in the far field the central peak of all five different four-slit combinations can be approximated by five times the first one, i.e., $ABCD$ (and accordingly for triple, double and single slits). Note that the following calculation and final result will be the same when treating those terms separately as we are only interested in the order of magnitude. The full treatment, however, would involve much more terms to track. For the central deviation, we can use that the ratio of the central peaks of the probability function of gratings is given by $G^{(1)}_{A_1,\ldots,A_N}(0)/G^{(1)}_{A_1,\ldots,A_N'}(0)=N/N'$ and we thus get
\begin{eqnarray}
\kappa^{(2)}(0) &\approx& 
2\Bigg(\frac{\Delta_{ABCDE} - 4 \Delta_{ABCD}+ 6\Delta_{ABC} - 4\Delta_{AB}+ \Delta_{A}}{G^{(1)}_{ABCDE}} \Bigg)  \, .
\end{eqnarray}
As in the main paper, we assume the deviations to be of the same (small) order of magnitude $\mathcal{O}(\Delta_X)=\Delta$ and by applying conventional propagation of uncertainty, we get $\kappa^{(2)}(0) \approx 2 \sqrt{16} / G^{(1)}_{ABCDE}(0)  \times \Delta$. Compared to the single-particle case, i.e., $\kappa^{(1)} \approx \sqrt{7}\Delta/G^{(1)}_{ABC}(0)$ (derived in the main paper), we get an improvement of $\kappa^{(2)}(0)/\kappa^{(1)} \approx 2 \cdot 3/5 \sqrt{16/7} \approx 1.8$, which is now independent of the original magnitude of the individual deviations $\Delta$.

For the $M$-particle case, one has to evaluate an accordingly extended expression of $\kappa^{(M)}(0)$. In the numerator, it comes down to counting the number of terms (while taking care of their prefactors), which involves the number of all comprised slit-combinations appearing in $I^{(M)}_{2M+1}$. This is given by the sum over all $k$-slit combinations with $1\leq k \leq 2M+1$, i.e.,
\begin{eqnarray}
\sum_{k=1}^{2M+1} \left| \mathcal{C}_{k}^{2M+1} \right| = \sum_{k=1}^{2M+1} \begin{pmatrix} 2M+1 \\ k \end{pmatrix} = 2^{2M+1}-1 \, ,
\end{eqnarray}
where $|.|$ stands for the cardinality. The sum runs from the $(2M+1)$ possible single-slits to the one possible $(2M+1)$-slit.
In the actual numerator of $\kappa^{(M)}(0)$, however, each slit-combination contributes with a different prefactor. Each addend is modified by the prefactor
\begin{eqnarray} 
\frac{G^{(M-1)}_{A_1,\ldots,A_{k}}(0)}{G^{(M-1)}_{A_1,\ldots,A_{2M+1}}(0)}=\left(\frac{k}{2M+1}\right)^{M-1} \, ,
\end{eqnarray}
where we already incorporated $M-1$ of the $M$ normalization factors of the denominator in the definition of the many-particle Sorkin parameter.
The determining number of contributing deviations is thus given by
\begin{eqnarray}
C(M) \vcentcolon=\sum_{k=1}^{2M+1} \left| \mathcal{C}_{k}^{2M+1} \right| \left(\frac{k}{2M+1}\right)^{M-1} = \sum_{k=1}^{2M+1} \begin{pmatrix} 2M+1 \\ k \end{pmatrix} \left(\frac{k}{2M+1}\right)^{M-1} \, ,
\end{eqnarray}
which can not be further simplified for arbitrary $M$.
Including the prefactor $M$ from Eq.~\eqref{eq:GM-deviations} and the remaining normalization factor $G^{(1)}_{A_1,\ldots,A_{2M+1}}(0)$, we finally can estimate the order of magnitude of the deviations in $M$-particle experiments, i.e.,
\begin{eqnarray}
\kappa^{(M)}(0) =  \frac{M \sqrt{C(M)} }{G^{(1)}_{A_1,\ldots,A_{2M+1}}(0)} \times \Delta + \mathcal{O}(\Delta^2)  \, ,
\end{eqnarray}
or normalized with respect to the single-particle case, 
\begin{eqnarray}\label{eq:increase}
\frac{\kappa^{(M)}(0)}{\kappa^{(1)}} = \frac{3 M }{2M+1}  \sqrt{ \frac{C(M) }{7}} = \frac{3 M }{(2M+1) \sqrt{7}} \left[ \sum_{k=1}^{2M+1} \begin{pmatrix} 2M+1 \\ k \end{pmatrix} \left(\frac{k}{2M+1}\right)^{M-1} \right]^{1/2} \, .
\end{eqnarray}
The last two equations indicate the increase in sensitivity of the family of many-particle Sorkin parameters to deviations from Born's rule. 
The numerical values of Eq.~\eqref{eq:increase} up to $(M=11)$-particle interference are 
\begin{center}
\begin{tabular}{ *{10}{c |} c}
	M & 2 & 3 & 4 & 5 & 6 & 7 & 8 & 9 & 10 & 11 \\ \hline		
	\vphantom{\Big(}$\kappa^{(M)}(0) / \kappa^{(1)} $ & 1.8 & 2.9 & 4.7 & 7.3 & 11.4 & 17.7 & 27.6 & 42.7 & 66.2 & 102.5
\end{tabular} \, . 
\end{center}